\documentclass[twocolumn,showpacs,preprintnumbers,amsmath,amssymb,showkeys]{revtex4-1}
\pdfoutput=1 

\usepackage{graphicx}% Include figure files
\usepackage{dcolumn}% Align table columns on decimal point
\usepackage{bm}% bold math
\usepackage{upgreek}
\usepackage{array}
\usepackage{booktabs}
\usepackage{microtype}
\usepackage{cmap}
\usepackage{geometry}
\usepackage[dvipsnames]{xcolor}

\usepackage{hyperref}
\hypersetup{colorlinks, citecolor={blue}, linkcolor={red}, urlcolor={violet}, pdftitle={A Kinetic Study of Amyloid Formation: Fibril Growth and Length Distributions}, pdfauthor={John S. Schreck, Jian-Min Yuan}, pdfdisplaydoctitle}

\begin{document}

\author{John S. Schreck}
\email{john.schreck@chem.ox.ac.uk}

\author{Jian-Min Yuan}
\email{yuanjm@drexel.edu}
\affiliation{Department of Physics, Drexel University, Philadelphia, PA 19104}

\title{A Kinetic Study of Amyloid Formation: Fibril Growth and Length Distributions}

\begin{abstract}
We propose a kinetic model for the self-aggregation by amyloid proteins. By extending several well-known models for protein aggregation, the time evolution of aggregate concentrations containing $r$ proteins, denoted $c_r(t)$, can be written in terms of generalized Smoluchowski kinetics. With this approach we take into account all possible aggregation and fragmentation reactions involving clusters of any size. Correspondingly, an aggregate of size x+y could be formed by or break-up into two smaller constituent aggregates of sizes x and y. The rates of each aggregation or fragmentation reaction, called kernels, are specified in terms of the aggregate size, and we solve $c_r(t)$ for large cluster sizes using numerical techniques. We show that by using Smoluchowski kinetics many pathways to fibrillation are possible and quantities, such as the aggregate length distribution at an arbitrary time, can be calculated. We show that the predicted results of the model are in agreement with the experimental observations.
\end{abstract}


\maketitle

\section{ Introduction }
Many experimental techniques have been developed to study and characterize the kinetic processes involved in the self-assembly of amyloid proteins into fibrils. The results of experiments involving amyloid formation {\it in vitro} have been used to formulate hypotheses about the pathological process, and treatment of diseases caused by amyloid proteins~\cite{Serio2000, Arvinte1993, Sokolowski2003, Hurshman2004, Frankenfield2009, Padrick2002, Chen2002, Modler2003, Collins2004, Ignatova2005}. However, experiments often report different results, especially during the early times in the aggregation process, before $\beta$-amyloid fibrils have formed. As a result, two competing hypotheses attempt to explain the results: the nucleated polymerization mechanism (NP), and the nucleated conformational conversion mechanism (NCC).  

Protein aggregation can be monitored by ThT or Congo Red dye, which may bind to aggregates that contain some amount of $\beta$-sheet content, mainly fibrils~\cite{Levine1999, Klunk1999, Naiki1999, Hartley1999, Nichols2005, Williams2005}. Thus, these dyes do not bind to oligomers that lack significant $\beta$-sheet content~\cite{Levine2008, Bieschke2005}. In experiments, the concentration of monomers when fibrils may form is referred to as the critical fibril concentration (CFC). The NP model contains several essential features: (1) the lag-time exists for a nucleus to form, and the concentrations of oligomers smaller than the nucleus are assumed to be negligible, (2) the concentration of amyloid protein monomers must exceed the critical fibril concentration for fibrils to form, and (3) the lag-time depends strongly on the initial amount of protein concentration present~\cite{Oosawa1975, Powers2006}. The equation for this relationship can be written as~\cite{Powers2006}
\begin{equation}
\label{scaling}
\ln t_{lag} = A - \left( \frac{ n_c + 1}{2} \right) \ln m_{tot}
\end{equation}
where $A$ is a constant, and $m_{tot}$ is the initial mass concentration of amyloid protein monomers. 

In contrast to the NP model, however, a variety of experiments~\cite{Lomakin1997, Garzon1997, Sabate2005, Ishii2007, Ono2009, Bernstein2009, Kelly2011} provide evidence for non-vanishing oligomer concentrations when the monomer concentration is below the CFC. These results led researchers to propose the NCC mechanism~\cite{Serio2000}, wherein quickly-forming oligomers undergo a slow conformational transition from a largely unstructured aggregate to a more organized nucleus that can grow into a $\beta$-sheet dominated fibril~\cite{Serio2000, Ishii2007, Kelly2011}. The same mechanism for nucleus formation has been proposed for other amyloid proteins including prions~\cite{Serio2000}, A$\beta$(1-40)~\cite{Kelly2011}, Huntington protein~\cite{Thakur2009}, and islet amyloid polypeptide~\cite{Wei2011}.  Lee, et al.~\cite{Kelly2011} has even recently suggested the NP mechanism ought to be abandoned for A$\beta$(1-40) in favor of the NCC mechanism. Since there may be major structural differences between the unstructured oligomers and the highly-ordered fibrils, some researchers have suggested that the early-time oligomers may resemble micelles, when the concentration of free monomers is above the critical micellar concentration (CMC)~\cite{Lomakin1997, lee, Schmit2011}. Thus, the process is characterized by a free-energy barrier separating unstructured oligomers, such as micelles, and structured fibril regimes~\cite{Ishii2007, Kelly2011}. 

Adding to these complications are the large number of possible aggregation and fragmentation reactions that determine whether fibrils grow longer, or whether they break up into smaller aggregates. For example, experiments and simulations have shown that the pathways to fibrils could be sequence dependent~\cite{Bitan2003, Urbanc2010}, and merging between aggregates of various sizes may play a role in the overall growth of fibrils. Kinetics models for amyloid formation often assume that the rates of elongation and fragmentation of aggregates have no size dependency, and are unrelated to any other properties that the amyloid proteins or aggregates may possess~\cite{Oosawa1975, Ferrone1985, Ferrone1999, Knowles2009, Morris2009}. The values of rate constants describing aggregation or fragmentation determine not only the mass of protein in fibrils as a function of time, but also the average lengths of the fibrils in time, two quantities that are readily measured in experiments~\cite{VanGestel2008, Xue2009, Adamcik2010, Morbidelli2012}. In other words, understanding which aggregation or fragmentation reactions contribute to the growth or shrinkage of fibrils, respectively, is necessary for accurately predicting quantities such as the total mass of protein in fibrils, the average lengths, and the size distributions of fibrils as a function of time. 

In the sections to follow, our aim is to introduce a kinetic model for amyloidogensis in which nuclei may form and undergo conformational transitions, where aggregates smaller than the critical size may exist in non-zero concentrations, and the length distributions of fibrils are accurately predicted. In Section ``Aggregate Systems Studied'', we consider in our description of fibril formation a variety of fibril elongation and fragmentation mechanisms, and then in Section ``Size-Dependent Kernels'', we show that the rates for aggregation or fragmentation steps should in general depend on the size of the aggregates. In Section ``Results'', the model predictions are compared with ThT and AFM experiments. Finally, in Section ``Generalized Kinetics'', we extend the kinetics approach introduced initially to include the kinetics of oligomers.

\section{Aggregate Systems Studied} 
\label{sec:one}

We consider a system containing a fixed initial concentration of monomeric amyloid proteins, $m_{tot}$, that are well mixed with solvent. The proteins are freely mobile in the solution and the system has a fixed volume, $V$. According to the law of mass-action, the reaction rate for reactions between any protein monomers or aggregates are proportional to the product of the reactant concentrations. In this mean-field description, we consider the possible aggregation pathways, from dimers, trimers, $\dots$, oligomers, $\dots$, proto-fibrils, up through fibrils. A monomer is defined as a protein that may or may not have a well-defined conformation, and can interact with surrounding solvent. However, we consider only interactions between proteins that may be responsible for the formation, elongation, or breaking of aggregates. An $N$-mer is an aggregate  containing $N$ monomers, which could range in size from dimers up to fibrils. Each monomer in an aggregate may interact with other neighboring proteins in the same aggregate via hydrophobic interactions or hydrogen bonds. 

\subsection{Critical nuclei} 

As we have discussed, there is much debate as to whether amyloid fibrils nucleate directly from protein solution, or whether oligomers form first and then convert into fibrils (or perhaps both processes are occurring simultaneously). It is commonly assumed in the NP kinetic models that the concentrations of aggregates that are smaller than the critical nucleus are negligible, and the critical nucleus represents the aggregate with the highest energy along the pathway. Since the NP and NCC mechanisms could be two extremes of some underlying nucleation mechanisms~\cite{Kelly2011, Ricchiuto2012}, we consider when a NCC model may become `NP-like', that is, it contains the feature that the concentration of nuclei is proportional to the monomer concentration raised to the $n_c$ power, as in the NP models of Oosawa-Kasai and Knowles, et al.,~\cite{Knowles2009} discussed below. The validity of the models discussed here are based on comparing the model predictions with experimental data, for example, data from ThT, AFM, and FlAsH fluorescence experiments. 

In this section, we consider a model in which a protein can exist in two different conformations, that we shall call A-type and B-type, respectively. The oligomers are assumed to have appreciable concentrations only as A-type aggregates, while the fibrils are of the B-type aggregates. The A-type aggregates could, for instance, resemble micelles and have a spherically symmetric shape~\cite{Lomakin1997, lee, Schmit2011}.  In NP kinetic models~\cite{Oosawa1975}, the critical nucleus forms via the interactions of $n_c$ number of proteins according to the reaction
\begin{equation}
\label{nuke_oosawa}
n_c C_1 \underset{k_2}{\overset{k_1}{\rightleftharpoons}} C_{n_c}^{A}
\end{equation}
where the concentrations of aggregates smaller than $n_c$ are assumed zero, and $k_1$ and $k_2$ are the formation and dissociation rate constants of the nucleus in solution. Once nuclei are formed, they may proceed to grow into fibrils. In our formulation, we incorporate a conformational transition of A-type to B-type aggregates of size $n_c$ as is illustrated in Fig.~\ref{pathwaysfig}, in accordance with a NCC mechanism. The reaction for conformationally induced transitions of A-type to B-type nuclei can be written as 
\begin{equation}
\label{conformation_nucleation}
C_{n_c}^A \underset{k^*}{\overset{k_n^{'}}{\rightleftharpoons}} C_{n_c}^B
\end{equation}
where $k_n^{'}$ and $k^*$ are, respectively, the rate constants for $n_c$ number of proteins converting from conformation A to B, and B to A reversibly~\cite{Pappu2011}. As soon as B-type aggregates form, they may grow into fibrils, where the fibril growth and shrinkage pathways are discussed in the next section. The rate of change of the B-type nuclei takes the form
\begin{equation} 
\frac{d c_{n_c}^B(t)}{d t} = k_n^{'} c_{n_c}^A(t) - k^* c_{n_c}^B(t).
\label{nuke_eq}
\end{equation}
In Eq. (\ref{nuke_eq}), there will also be the rates where $c_{n_c}^B$ is converted to larger oligomers. These pathways are discussed in the next section. If the A-type micelles form rapidly and convert into B-types, the A-type oligomer phase can be effectively by-passed~\cite{Auer2012}. The reactions in Eq.~(\ref{nuke_oosawa}) are taken to be in kinetic equilibrium, thus Eq.~(\ref{nuke_eq}) can be written as
\begin{equation} 
\frac{d c_{n_c}^B(t)}{d t} = k_n c_1(t)^{n_c} - k^* c_{n_c}^B(t)
\label{nuke_eq2}
\end{equation}
which is similar to the equation used by Oosawa and Asakura~\cite{Oosawa1975}, and others~\cite{Knowles2009}, for nuclei production of constant size and we have defined $k_n \equiv k_1 k_n^{'} / k_2$. In Section ``Generalized Kinetics'' we will consider Eq.~(\ref{nuke_eq}) for the case when oligomers may reach metastable states instead of directly going to equilibrium, and can compare this approach to the formation of fibrils with the FlAsH fluorescence experiments carried out by Lee, et al.~\cite{Kelly2011}. 

In the above treatment of the fibril nucleus, we assumed that the initial monomeric protein solution was highly super-saturated and that a pre-equilibrium between monomers in solution and the fibril nucleus of size $n_c$ exists. However, as more and more monomers join aggregates, the solution eventually becomes only weakly super-saturated or saturated so that new fibrils can only form from either the breaking of larger aggregates or through nucleation events. In these cases, the size of the fibril nucleus may not be constant or correspond to a single $n_c$ value and Eq.~(\ref{nuke_eq2}) is not strictly justified. Furthermore, Kashchiev and co-workers have pointed out that the fibril nucleation rates cannot be described by Eq.~(\ref{nuke_eq2}), even if the nucleus size remained constant. These points are well discussed in their recent paper~\cite{Kashchiev2013}. We nevertheless use Eq.~(\ref{nuke_eq2}) because it is well known to reproduce the scaling relation of Eq.~(\ref{scaling}) when comparing the model to ThT experiments. In future studies of protein amyloid aggregates involving mass-action equations, we hope to address the peculiar problem of fibril nucleation more correctly. To begin, we first focus on using the NP-like description to study how fibrils may grow and shrink in size, which are compared with ThT and AFM experiments. This is discussed in the next section. 

\subsection{Fibril formation pathways}
%%%%%%%%%%%%%%%%%%%%%%%%%%%%%%%%%%%%%%%%
	\begin{figure}[t]
	\begin{center}
	\vspace{0.6 cm}
	\includegraphics[width = 245 pt  ]{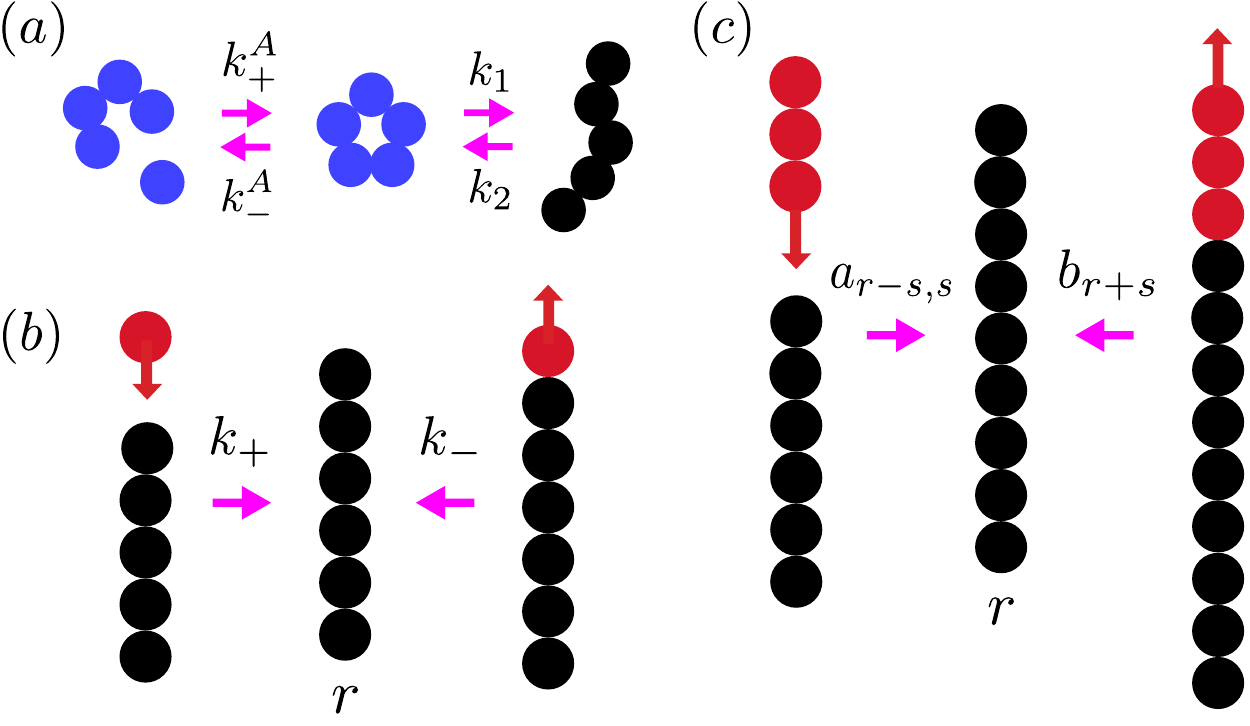}
	\caption{(a) A small aggregate of size $n_c$ could undergo a conformational transition from a structured (blue) to an aggregate which contains $\beta$-sheet (black). (b) The Becker-D{\"o}ring type kinetics of monomer addition and subtraction pathways. (c) Illustrating the Smoluchowski aggregation and fragmentation pathways. The Becker-D{\"o}ring kinetics is a special case of Smoluchowski kinetics. All illustrations are schematic; the aggregates in (c), for example, could merge laterally. }
	\label{pathwaysfig}
	\end{center}
	\end{figure}
%%%%%%%%%%%%%%%%%%%%%%%%%%%%%%%%%%%%%%%%

In their model for actin polymerization, Oosawa and Asakura assumed the pathways for actin filaments to grow or shrink in size are monomer addition and monomer subtraction, respectively~\cite{Oosawa1975}. These reactions are just two of the many possible pathways for amyloid aggregate growth and shrinkage. While it could be the case that these two reactions dominate the changes in aggregate sizes, other types of growth and fragmentation mechanisms may also contribute. For example, Knowles et al.~\cite{Knowles2009}, and others~\cite{Hong2011} have showed that strong fragmentation can act as an effective secondary nucleation mechanism, whereby fragment aggregates can act as seeds that speed up elongation of fibrils during early times of aggregation. For the case of strong fragmentation, $t_{1/2} \sim m_{tot}^{-1/2}$, otherwise $t_{1/2} \sim m_{tot}^{-(n_c +1)/2}$, which is the prediction of the Oosawa-Asakura NP model.

By generalizing the Becker-D{\"o}ring approach, that is, monomer addition and subtraction, all possible ways of forming an aggregate containing $r$ proteins can be taken into account~\cite{Schreck2011}. In the most general case, any two fibrils of sizes $r$ and $s$ could merge to form a larger aggregate, which can be described the rate constant $a_{r,s}$. Alternatively, a fibril of size $r+s$ could break into two pieces of sizes $r$ and $s$, respectively, which can be described by the rate constant $b_{r+s}$. Both of these scenarios are illustrated in Fig.~\ref{pathwaysfig}(c). The expression for these types of reactions can be written as
\begin{equation}
\label{smol2}
C_r+ C_s \underset{b_{r+s}}{\overset{a_{r,s}}{\rightleftharpoons}} C_{r+s}
\end{equation}
and is referred to as generalized Smoluchowski kinetics~\cite{Smol1916, Wattis2006}. The aggregation and fragmentation rate constants $a_{r,s}$ and $b_{r+s}$, respectively, are referred to as kernels and must be symmetric, that is $a_{r,s}=a_{s,r}$ and $b_{r+s} = b_{s+r}$. By imposing the law of mass action, the mass flux from aggregate concentrations $c_{r}(t)$ and $c_{s}(t)$ going to $c_{r+s}(t)$ can be written as 
\begin{equation}
\label{smolflux2}
W_{r,s}(t) = a_{r,s} c_{r}(t)c_{s}(t) - b_{r+s} c_{r+s}(t)
\end{equation}
where the NP nucleation phenomena described by Eq.~(\ref{nuke_eq}) restricts $c_r(t)=0$ if $r < n_c$ except for $r=1$. If $n_c \ge 2$, $W_{1,1}(t)=0$. The corresponding mass-action equations for the evolution of an aggregate of size $r$, where $r\ge n_c$, can be written using Eq.~(\ref{smolflux2}) as 
\begin{equation} \begin{split}
\label{smoluchowski}
\frac{d c_r(t)}{dt} &= \frac{1}{2} \sum_{s=1}^{r-1} W_{s,r-s}(t) - \sum_{s=1}^{\infty} W_{r,s}(t) \\ &\quad +\left(k_{n} c_{1}(t)^{n_c} - k^* c_{n_c}(t) \right)\delta_{r,n_c}
\end{split} \end{equation} 
where the first term in Eq.~(\ref{smoluchowski}) counts every way to construct the aggregate of size $r$ twice, hence the factor of one-half. The first two terms describe all possible ways to for the aggregate of size $r$ to form and to disintegrate. The rate constants $k_n$ and $k^*$ were defined above, and the Kronecker delta equals one if $r=n_c$, and zero otherwise. To reduce the total number of parameters in Eq.~(\ref{smoluchowski}), we assume oligomers can only convert from A- to B-type, and that $k^*$ may be set to zero. Knowles' model combines monomer addition and subtraction, with the fragmentation part of reactions of the generalized Smoluchowski model~\cite{Masel1999, Knowles2009}. Since Knowles' model is well-studied~\cite{ Cohen2011a, Cohen2011b, Cohen2011c}, we compare with it our results obtained for the Smoluchowski model of protein aggregation. The resulting equations for the evolution of cluster sizes $c_r(t)$ in Knowles' model are
\begin{equation}
\label{knowles} \begin{split}
\frac{d c_r(t)}{dt} &= 2 k_{+} c_1(t) \left[ c_{r-1}(t) - c_{r}(t) \right] - k_{-} (r-1) c_{r}(t) \\ & \quad + 2 k_{-} \sum_{i=r+1}^{\infty} c_{i}(t) + k_n c_1(t)^{n_c} \delta_{r,n_c} 
\end{split} \end{equation}
where $k_+$ is the monomer addition rate constant, $k_{-}$ is the rate constant of any type of breaking up of an aggregate into two pieces, regardless of the sizes of the fragments, and $k_n$ describes the nucleus formation in Eq.~(\ref{knowles}). From the conservation of mass condition, $m_{tot} =  \sum_{k=1}^{\infty} k c_k(t)$, the evolution of the monomer concentration, $c_1(t)$, in Eqs.~(\ref{smoluchowski}) and (\ref{knowles}) can be written as
\begin{equation}
\label{monomer}
\frac{d c_1(t)}{dt} = - \sum_{j=n_c}^{\infty} j \frac{d c_j(t)}{dt} .
\end{equation}
In Eqs.~(\ref{smoluchowski}) and (\ref{knowles}), assuming $a_{r,s}=k_+$ and $b_{r+s}=k_-$ can be a very useful approximation that make these equations easier to solve, and also the number of parameters needed to capture the aggregation and fragmentation processes can be kept to a minimum. However, in reality the physics should play a role in determining $a_{r,s}$ and $b_{r+s}$. The size, shape, and even the conformation of monomers and aggregates, as well as thermal breaking~\cite{Lee2009thermal}, may influence the rates. With Becker-D{\"o}ring and Smoluchowski kinetics, we can consider the long-time behavior of the aggregate concentrations, $c_r(t)$, and derive expressions relating the fragmentation rates with size-dependent aggregation rates. In the next section, we show that once an aggregation kernel is determined, the fragmentation kernel can also be determined, and Eq.~(\ref{smoluchowski}) is solvable for size dependent kernels. 

\section{Size-Dependent Kernels} 
Two asymptotic states of the system described by Eqs. (\ref{smoluchowski}) and (\ref{knowles}) are usually of interest:
\begin{enumerate}
\label{steadystate}
\item The system reaches a {\it steady state}, when all the fluxes are constant as defined in Eq. (\ref{smolflux2}), all of the $c_r(t)$'s do not change in time, that is, $W_{r,s}(t) = W$ for all $r$ and $s$, where $W$ is a constant flux. 
\label{equil}
\item The system reaches {\it equilibrium}, where in Eq. (\ref{smolflux2}) the forward reaction exactly balances the backwards reaction, i.e., $W_{r,s}(t)=0$ for all $r$ and $t$. 
\end{enumerate}

We assume that our protein systems will eventually tend toward thermodynamic equilibrium and invoke detailed balancing, i.e., case 2 above, where the Smoluchowski mass flux, Eq.~(\ref{smolflux2}), equals zero for all $r$ and $s$. Thus, we may write Eq.~(\ref{smolflux2}) as 
\begin{equation}
\label{bd_balance}
a_{r,s} \rho_{r} \rho_{s} = b_{r+s} \rho_{r+s}
\end{equation}
where $\rho_i$ is the equilibrium concentration for the aggregate containing $i$ monomers and the equilibrium dissociation constant $K=\rho_{r} \rho_{s} / \rho_{r+s} = b_{r+s}/a_{r,s}$. The equilibrium concentrations $\rho_r$ can be expressed as $\rho_r = Z_r / V$ where $Z_r$ is the partition function of an aggregate of size $r$~\cite{Abraham1974, Wattis2006, Ferrone2006} and $V$ is the volume of the system. Solving Eq.~(\ref{bd_balance}) for $b_{r+s}$ yields
\begin{equation}
\label{eqconst}
b_{r+s} = \frac{ Z_r  Z_s } { Z_{r+s} } a_{r,s}.
\end{equation}
As an example, $Z_N$ for an $N$-mer aggregate can be written as~\cite{Abraham1974} 
\begin{equation}
Z_N = Z_{trans} Q_{in}
\end{equation}
where $Z_{trans}$ is the contribution from translational motion of the aggregate, and $Q_{in}$ is the contribution due to internal motion and interactions between proteins in the aggregate. $Q_{in}$ contains the contributions from rotational degrees of freedom for proteins in the aggregate, as well as collective vibrations of the aggregate itself. If the aggregates are regarded as being linear chains, the contribution due to the vibrations has a simple dependence on $N$~\cite{Abraham1974, Hill1980, Hill2002, Hill2002a} and $Z_{N}$ can be written as~\cite{Hill1983}
\begin{equation}
\label{pfapprox}
Z_N=X q^{N} N^{n} V
\end{equation}
where $X$ is a composite factor, $q^{N}V$ is the contribution of an infinite aggregate, and $X N^n$ takes into account the boundary effects due to finitely-sized aggregates. The parameter $n$ depends on the exact description of the proteins in aggregates as well as the collective properties and typically ranges between 4 and 6~\cite{Hill1983}. In this article, we assume $n=4$ in all calculations. Plugging Eq.~(\ref{pfapprox}) into Eq.~(\ref{eqconst}) and simplifying yields
\begin{eqnarray}
\label{bfroma}
b_{r+s}  &=& X \left( \frac{r s}{r+s} \right)^n   a_{r,s}
\end{eqnarray}
and only the problem of determining the aggregation kernel $a_{r,s}$ is left. 

\subsection{Protein Diffusion}

%%%%%%%%%%%%%%%%%%%%%%%%%%%%%%%%%%%%%%%%
	\begin{figure}[t]
	\begin{center}
	\includegraphics[width = 245 pt]{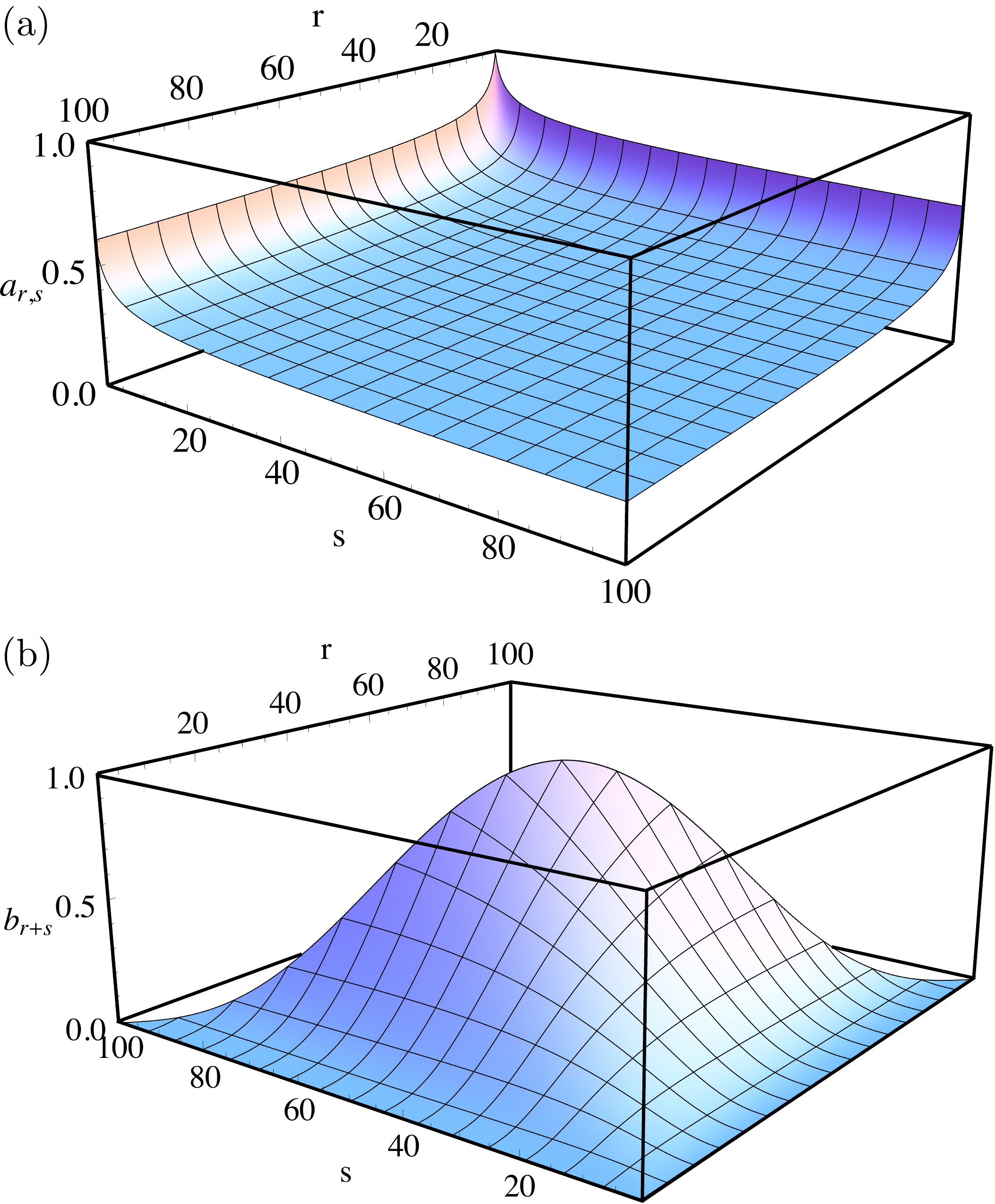}
	\caption{Illustrating Eq.~(\ref{sphere_kernel}) in (a), and Eq.~(\ref{frac_kernel}) in (b). In both cases the kernels were normalized for illustrative purposes. The plots show that both kernels exhibit symmetry in indices $r$ and $s$. In (b), the cut off size for aggregates was set to 100, meaning $r+s \le 100$. Additionally $D_f=3$ was used in Eqs.~(\ref{sphere_kernel}) and (\ref{frac_kernel}).}
	\label{aggkernels}
	\end{center}
	\end{figure}
%%%%%%%%%%%%%%%%%%%%%%%%%%%%%%%%%%%%%%%%

If the motions of monomers and aggregates in solvent are assumed to be controlled by diffusion alone, results from coagulation theory~\cite{Carrotta2005, Chandrasekhar1943} may be used to write the aggregation rates as 
\begin{equation}
\label{agg_rate}
a_{i,j} \propto \left( D_i + D_j \right) \left(R_i +R_j \right)
\end{equation}
where $D_{i}$ is the diffusion coefficient for a cluster of radius $i$, and is equal to $D_i = k_{B} T / 6 \pi i \eta$ where $\eta$ is the viscosity of the solvent. $R_i$ is the radius of the sphere of influence of the aggregate of radius $i$, which represents the maximum spatial separation two aggregates could be from each other and still stick together. In absence of understanding how the structure of aggregates might relate to reactivity, Eq.~(\ref{agg_rate}) is simplified and compared with previous studies~\cite{Hill1983} of aggregation and fragmentation of linear aggregates by assuming that $R_{i}$ is on the order of the aggregate's hydrodynamic radius~\cite{Carrotta2005}. A simplified expression for $a_{i,j}$ can be re-expressed in terms of $r$, the number of proteins in the aggregate, and the fractal dimension of the aggregate, $D_f$, as~\cite{Feder1984, Arosio2012}
\begin{eqnarray}
a_{r,s} &=& k_P \left( r^{-1/D_{f}} + s^{-1/D_{f}} \right)
\label{sphere_kernel}
\end{eqnarray}
where $k_P$ is a composite factor. The fractal dimension can be thought of as a measure of the compactness of the aggregate. For spherical monomers and aggregates $D_f=3$. This value would probably be an oversimplification in our model because fibrils are not spherical entities, rather they are linear, non-branching structures. A value of $D_f\approx2.56$ has been found for antibody aggregates~\cite{Feder1984, Arosio2012}. It is important to note that many different forms of kernels have been used in the literature for studying protein aggregation~\cite{Chandrasekhar1943, Hill1980, Hill1983, DaCosta1998, Pallitto2001, Carrotta2005, Lee2007, Hall2009, Ghosh2010, Hong2013}, and that other kernels may also be valid for various kinetics scenarios. From Eqs.~(\ref{bfroma}) and (\ref{sphere_kernel}), the fragmentation kernel can be written as 
\begin{eqnarray}
\label{frac_kernel}
b_{r+s} &=& k_{M} \left( \frac{r s}{r+s} \right)^n \left(r^{-1/D_{f}} + s^{-1/D_{f}} \right),  
\end{eqnarray}
where $k_M$ is also a composite factor. In the next section, we use $k_P$, $k_M$, $k_n$, and $D_f$ as adjustable parameters when comparing with experiments.

The aggregation kernel, Eq.~(\ref{sphere_kernel}), is plotted in Fig.~\ref{aggkernels}(a). As illustrated, monomer addition is predicted to be the favored pathway for small aggregates to grow larger, but this elongation mechanism weakens as the system size grows larger. Additionally, the aggregation kernel describes the merging of any two aggregates, where smaller aggregates are more likely to merge with larger ones than larger with larger. When aggregates grow large in size, the overall effect of merging on aggregate size progressively becomes weaker. 

In Fig.~\ref{aggkernels}(b), $b_{r+s}$ appears Gaussian and reaches a maximum when both $r$ and $s$ reach one-half the cut-off size for a finite system. The figure clearly shows that aggregates are more likely to break in the middle, and they become even more likely to break as overall aggregate sizes grow larger. Monomer subtraction is the least likely scenario for aggregate shrinkage. Additionally, increasing $n$ in Eq.~(\ref{frac_kernel}) (not shown) has the effect of reducing the overall ability of the aggregate to break into any two pieces, while progressively favoring breaking nearer to the center of the aggregate. Overall, the aggregation and fragmentation kernels yield similar predictions as Hill's studies of linear aggregates~\cite{Hill1983}. 

\section{Results} 
In this section, Eqs.~(\ref{smoluchowski}) and (\ref{knowles}) are solved together with Eq.~(\ref{monomer}) in each case, and some quantities that can be compared with experiments are obtained. We first study the constant kernel formulations for both the Knowles and Smoluchowski models, then consider size-dependent kernels given by Eqs.~(\ref{sphere_kernel}) and (\ref{frac_kernel}) for the Smoluchowski model. In order to compare with experiments, the moment generating function, $M_n(t)$, for the set $\{ c_{k}(t) \}^{\infty}_{k=1}$ is defined as $M_n(t) \equiv \sum_{k=n_c}^{\infty} k^n c_k(t)$ so that the number of proteins in aggregates, $P(t) = M_0(t)$, and the mass of aggregates, $M(t) = M_1(t)$, are given by 
\begin{eqnarray}
\label{number}
P(t) &=&  \sum_{k=n_c}^{\infty} c_k(t) \\ 
\label{mass}
M(t) &=& \sum_{k=n_c}^{\infty} k c_k(t)
\end{eqnarray}
and the average length of aggregates, $L(t)$, can be written as 
\begin{equation}
\label{length}
L(t) = \frac{M(t)}{P(t)}.
\end{equation}
Other quantities, such as the fluctuation in the number of proteins in aggregates, are easily derived by using the moment generating function $M_n(t)$. 

Eq.~(\ref{knowles}) is often solved in terms of $P(t)$ and $M(t)$, where in some cases exact expressions are attainable. In our approach, we solve the equations for aggregate concentrations described by Eqs.~(\ref{smoluchowski}) and (\ref{knowles}) numerically. Numerical solutions require that a cut-off size be introduced so that integration of the coupled set of differential equations in both models is tractable. Thus, in Knowles' model and the Smoluchowski model, the maximum size of an aggregate is denoted $\lambda_b$. Conserving mass requires us to truncate Eqs.~(\ref{smoluchowski}) and (\ref{knowles}) so that mass-loss cannot occur, therefore we impose the following condition to the set of concentrations $\{ c_{k}(t) \}^{\infty}_{k=1}$ in each model for all times: $W_{r,s}(t) = 0$ if $r+s > \lambda_b$ in Eq. (\ref{smoluchowski}), and $c_r(t) = 0$ if $r+1 > \lambda_b$ in Eq. (\ref{knowles}). Once the set of concentrations $\{ c_{k}(t) \}^{\infty}_{k=1}$ is computed, the mass and lengths of aggregates can be calculated.

\subsection{Constant kernel results}

%%%%%%%%%%%%%%%%%%%%%%%%%%%%%%%%%%%%%%%%
	\begin{figure}[t]
	\begin{center}
	\includegraphics[width = 245 pt]{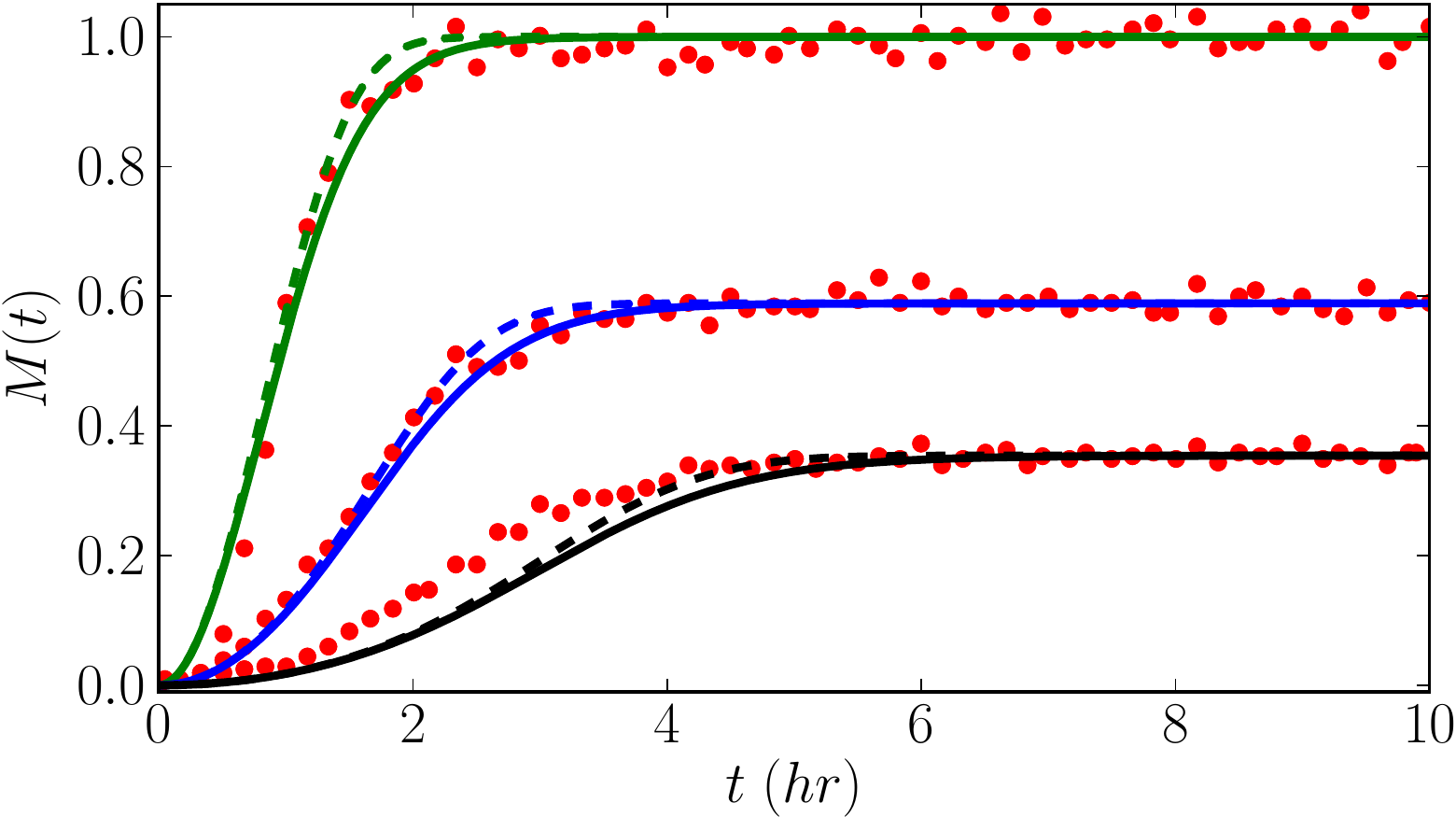} ~
	\caption{ The mass in aggregates $M(t)$ in Knowles' model (dashed lines) is fitted to the results of Hammer et. al, for three different initial mass concentrations, $m_{tot}$, of CgsB: 70, 35, and 17.5 $\mu$M from top to bottom, respectively. $M(t)$ for the Smoluchowski model is plotted (solid lines) and was fit to the same experimental data. The parameters used in Knowles' model were $k_{+}=2.65\cdot10^{4}$ $\text{M}^{-1} \text{s}^{-1}$, $k_{-}=9\cdot10^{-8}~\text{s}^{-1}$. In the Smoluchowski model, the parameters used were $k_{+}=5.3\cdot10^{4}$ $\text{M}^{-1} \text{s}^{-1}$, $k_{-}=1.8\cdot10^{-7}~\text{s}^{-1}$. In both models $k_n=4.62\cdot10^{-4}$ $\text{M}^{-1} \text{s}^{-1}$ was used. Experimental data points are shown as red dots in the figure. }
	\label{mass_plot_cgsa}
	\end{center}
	\end{figure}
%%%%%%%%%%%%%%%%%%%%%%%%%%%%%%%%%%%%%%%%
%%%%%%%%%%%%%%%%%%%%%%%%%%%%%%%%%%%%%%%%
	\begin{figure*}[t]
	\begin{center}
	\begin{tabular}{cc}
	\includegraphics[width = 200 pt]{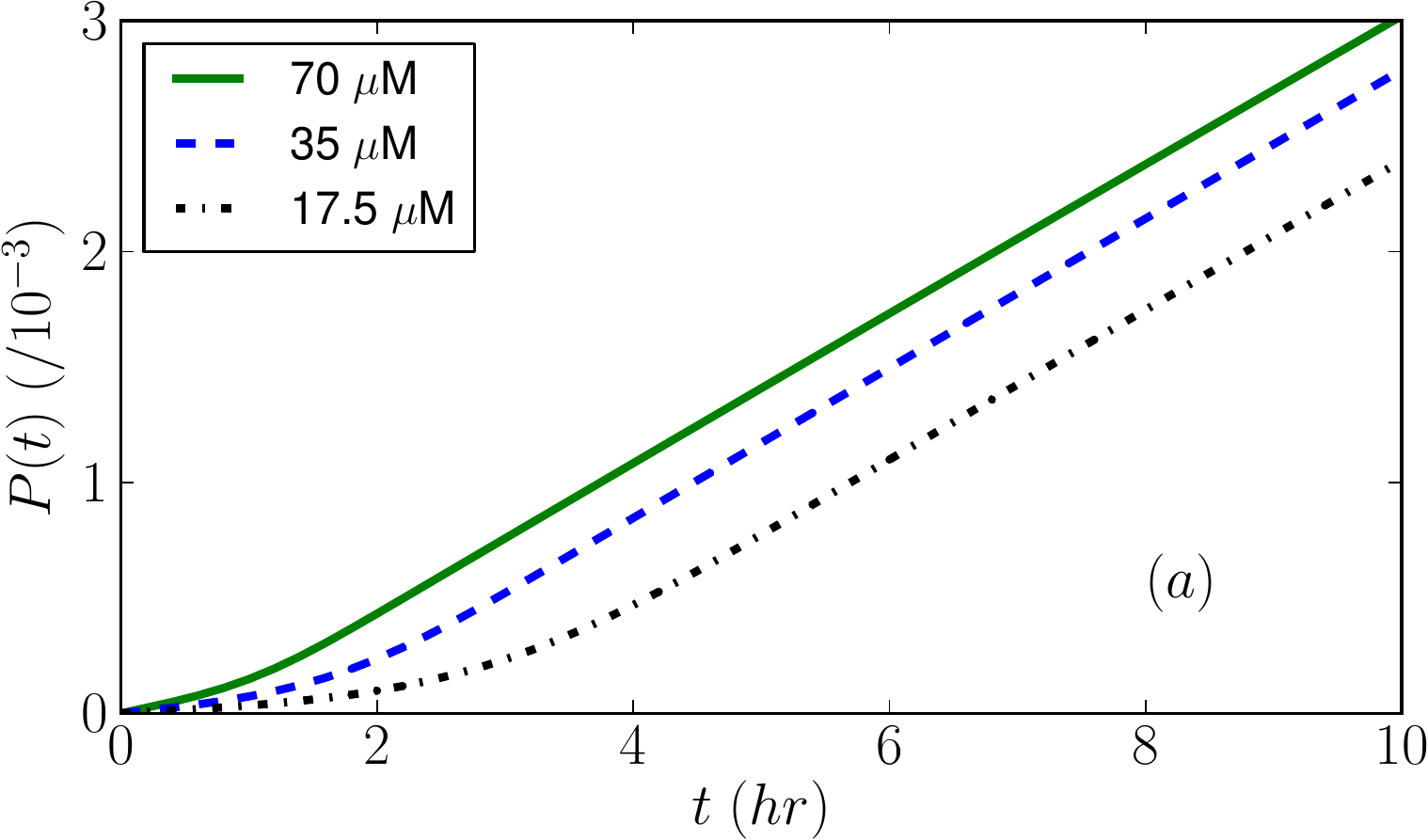}  & \includegraphics[width = 200 pt]{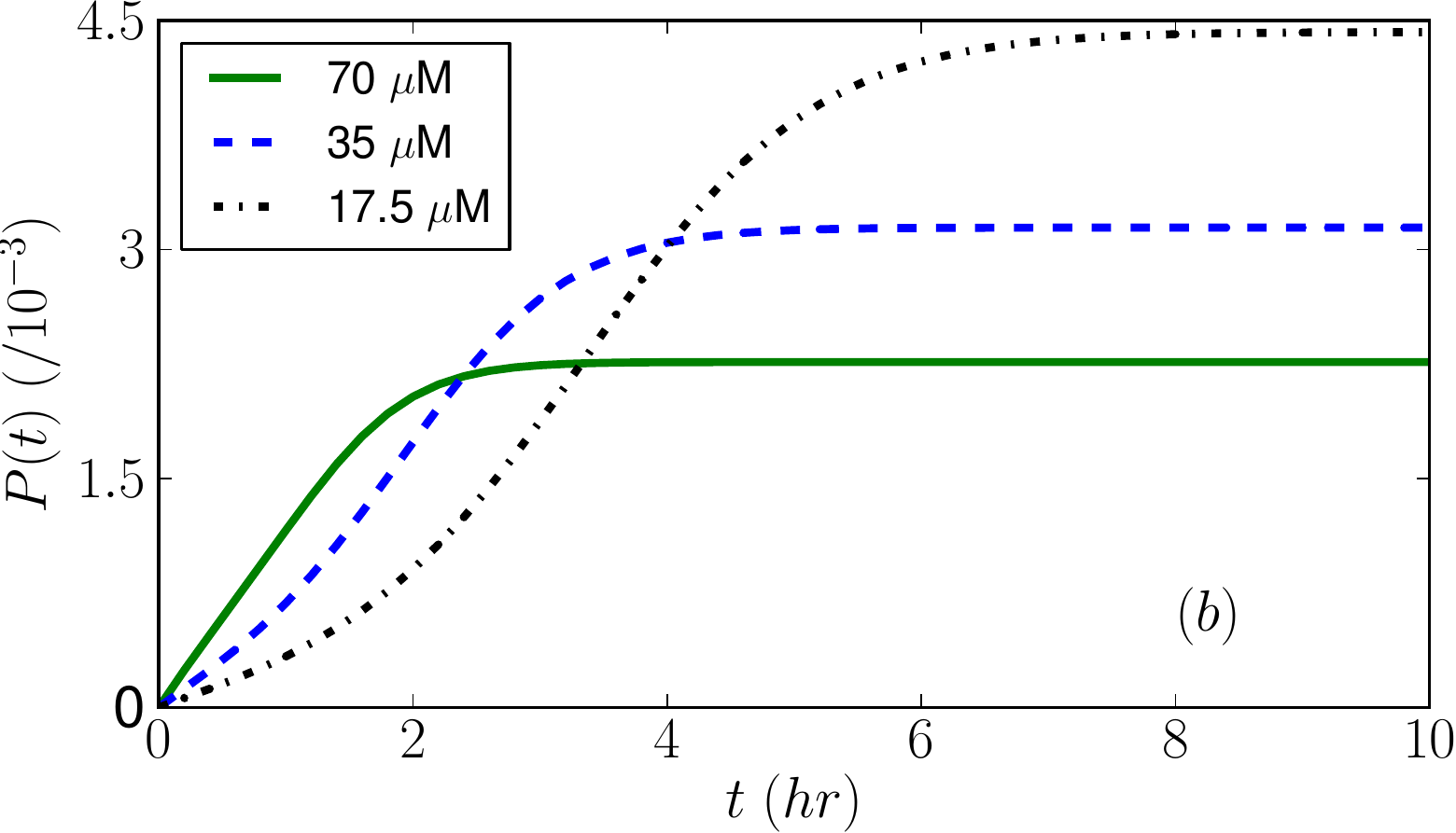} \\
	\includegraphics[width = 200 pt]{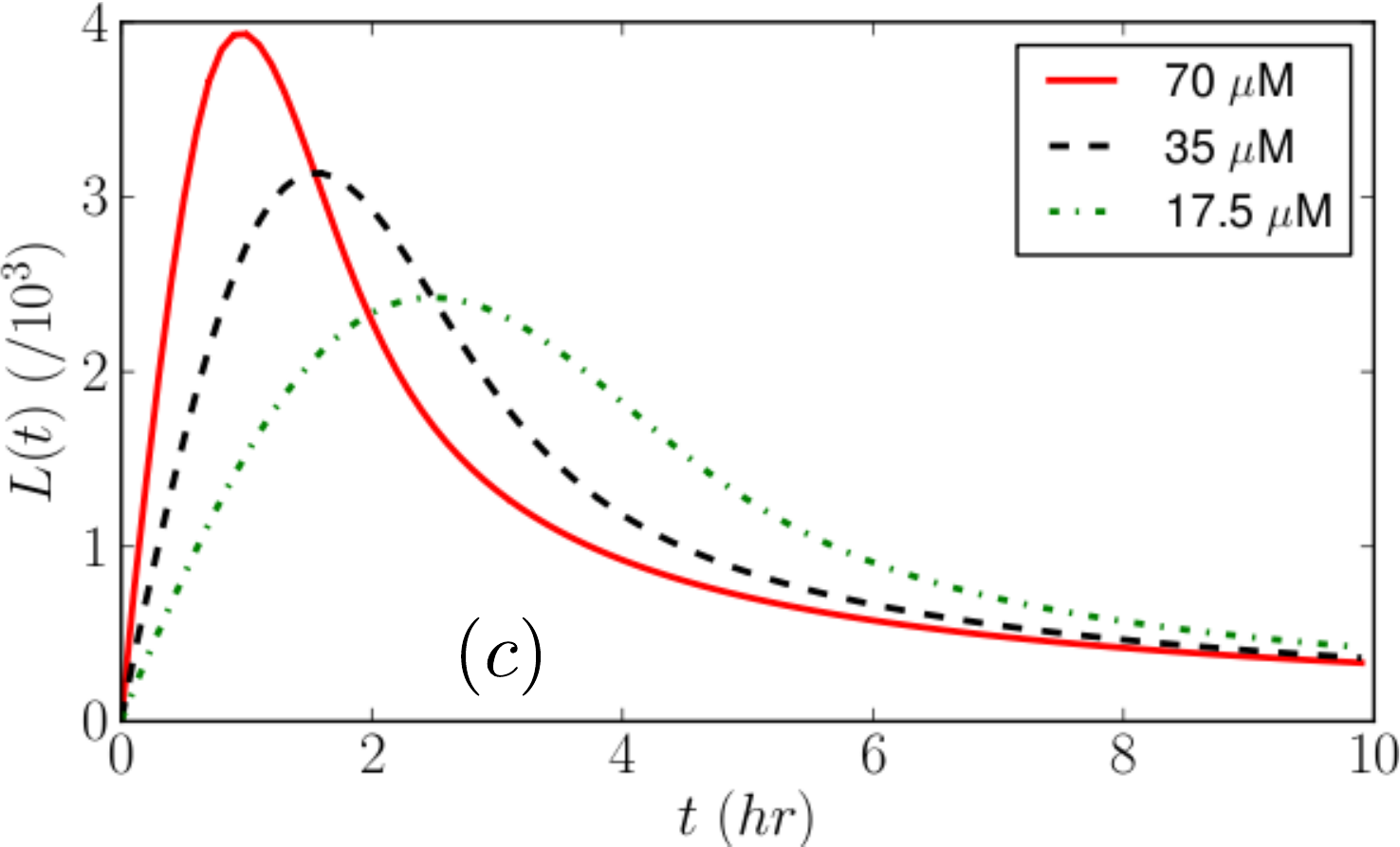}  & \includegraphics[width = 200 pt]{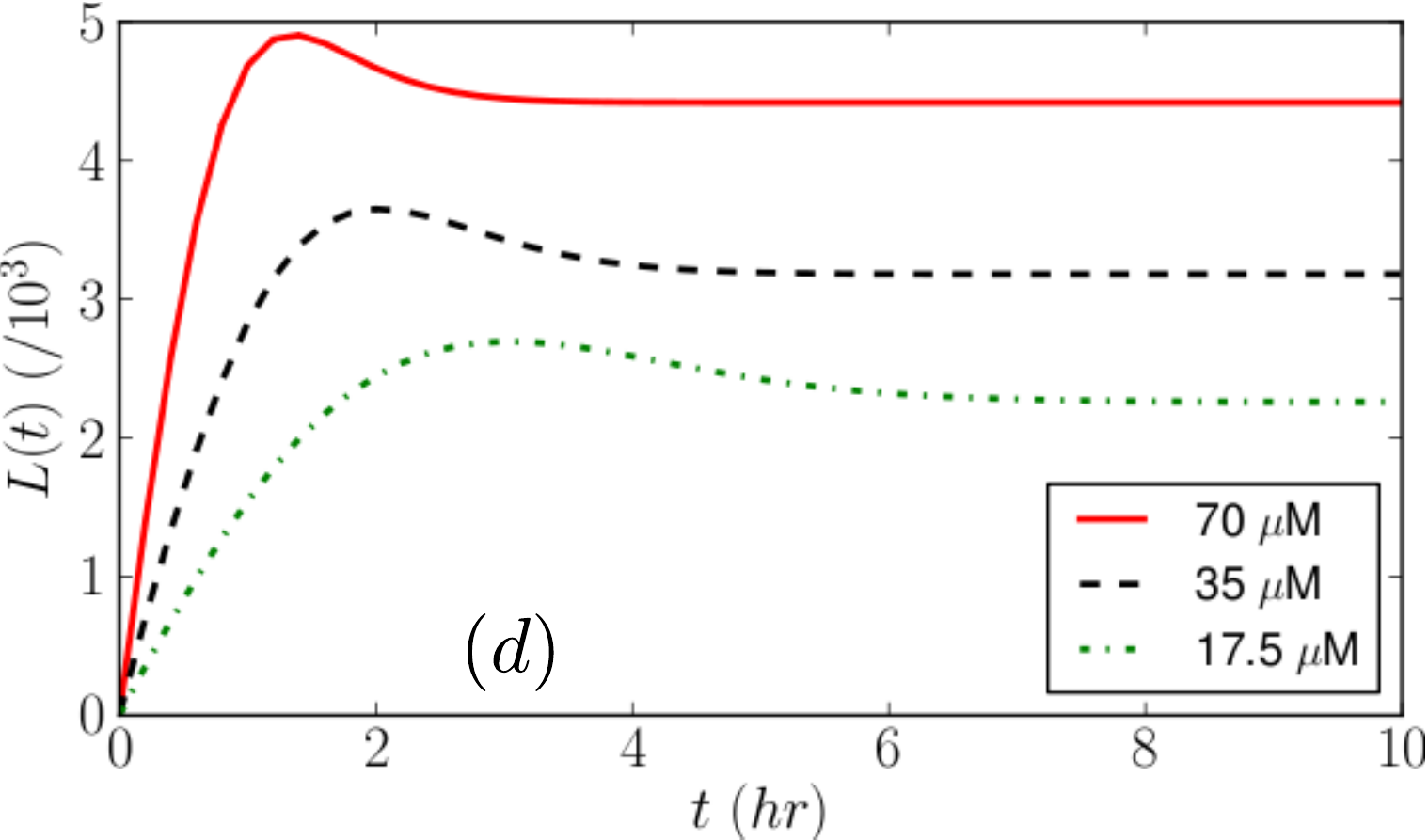} 
	\end{tabular}
	\caption{ In (a) the number of aggregates $P(t)$ for Knowles' model is shown using the same fit parameters as in Fig.~\ref{mass_plot_cgsa}(a) for the three different initial mass concentrations of CgsB: 70, 35, and 17.5 $\mu$M from top to bottom, respectively. In (b) we plot $P(t)$ for the Smoluchowski model using the same fit parameters as in Fig.~\ref{mass_plot_cgsa}(b). In (c) the average length of fibrils, $L(t)$, is plotted for Knowles' model, while in (d) $L(t)$ is plotted for the Smoluchowski model.  }
	\label{numb_plot}
	\end{center}
	\end{figure*}
%%%%%%%%%%%%%%%%%%%%%%%%%%%%%%%%%%%%%%%%

In Fig.~\ref{mass_plot_cgsa}, we fit the mass in fibrils, $M(t)$, to the 70$\mu$M ThT fluorescence data set for Curli fibrils by adjusting $k_{+}$, $k_{-}$, and $k_n$ in Eqs.~(\ref{smoluchowski}) and (\ref{knowles})~\cite{Hammer2012}. The fit parameters where then used to predict $M(t)$ for the 35$\mu$M and 17.5$\mu$M data sets. The fluorescence signal was scaled in accordance with the initial amount of protein used. Additionally, the critical nucleus size for Curli fibrils is regarded as $n_c=2$~\cite{hammer, Hong2011, Hammer2012}. Both the Smoluchowski and Knowles' model fit the 70$\mu$M and 35$\mu$M data sets quite well, but the fits are not as good for the 17.5$\mu$M data set. Overall, Knowles' model  has been shown to reliably fit a variety of ThT data for many different proteins~\cite{Knowles2009, Hong2011} with different critical nucleus sizes. However, the Smoluchowski model has not been as extensively studied for protein aggregation, except for a few cases ~\cite{Carrotta2005, Arosio2012, Hong2013}. 

It is not clear from the mass plots in Fig.~\ref{mass_plot_cgsa} that there are significant differences between the predictions of the Knowles and Smoluchowski models. However, since the ThT signal is proportional to the mass of protein in fibrils, and not the fibril sizes, a flattening in the ThT signal does not imply the fibrils have necessarily stopped growing longer or shorter. In other words, just fitting the mass of fibrils, $M(t)$, to ThT curves is not enough to establish which model, if any, is the better model for protein aggregation. Thus, we turn our attention to the time evolutions of the number of aggregates and the aggregate average lengths. In Fig.~\ref{numb_plot}, we plot the total number of aggregates, $P(t)$, for both Smoluchowski and Knowles models using the fits of the ThT curves for Curli fibrils. For all cases of initial mass concentration, the number of aggregates increases from zero and increases almost linearly. The trend in Knowles' model  seems to be for higher initial mass concentrations, more aggregates are present at all times studied. The Smoluchowski model predicts significantly different behavior. For smaller mass concentrations, there are more aggregates present at early times, but as time progresses, these roles seem to reverse: there are more aggregates for higher initial mass concentrations when compared to lower mass concentrations. Interestingly, as the initial mass increases in the Smoluchowski model, the number of aggregates, $P(t)$, exhibits a sigmoidal-like curve, in contrast to the effectively linear growth in the number of fibrils in Knowles' model  at higher times.  This seems to suggest that the average lengths in the Smoluchowski model will converge towards a constant value, whereas in Knowles' model  the average lengths will become shorter for the time scales studied here. 

%%%%%%%%%%%%%%%%%%%%%%%%%%%%%%%%%%%%%%%%
	\begin{figure*}[t]
	\begin{center}
	\begin{tabular}{cc}
	\includegraphics[width = 200 pt ]{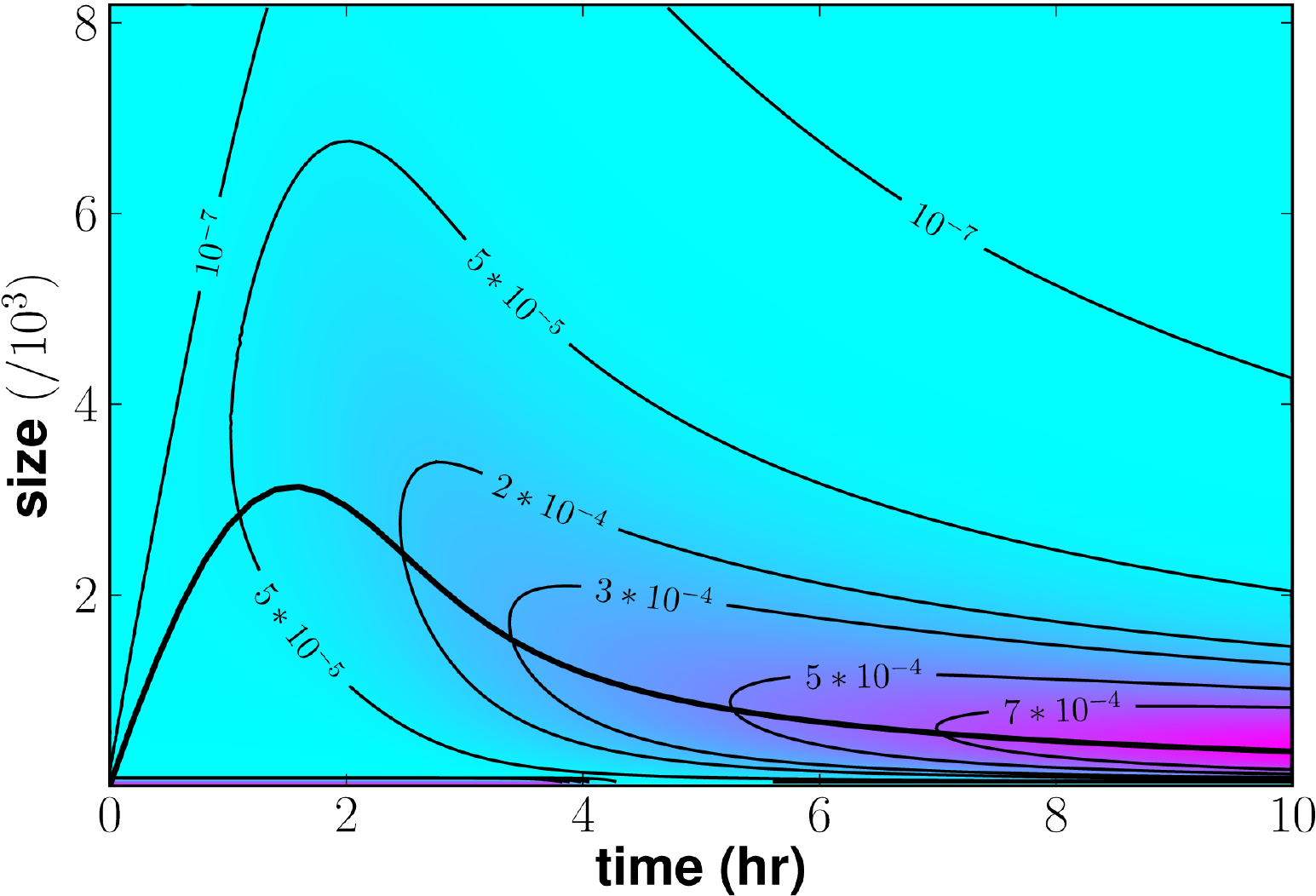} & \includegraphics[width = 200 pt]{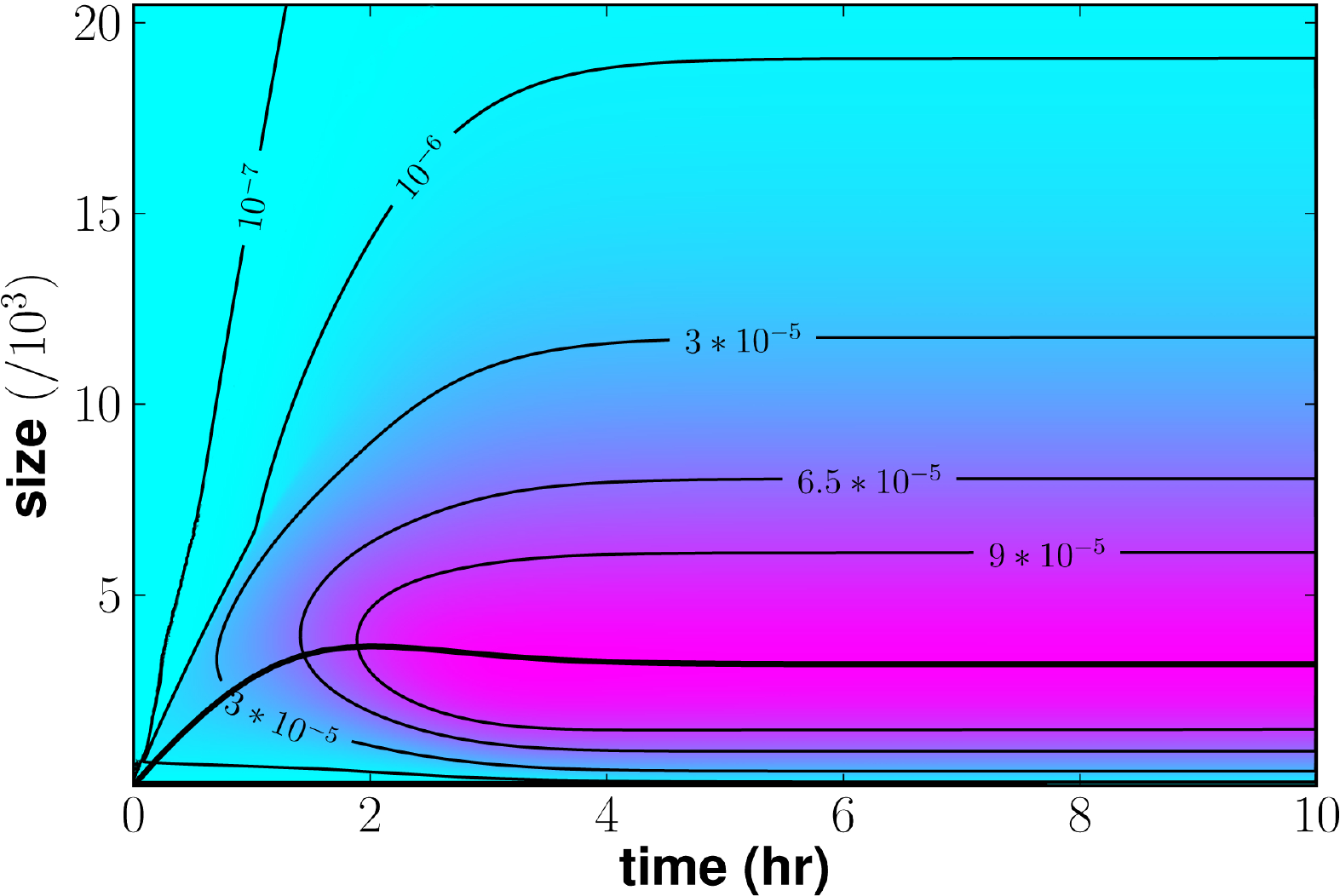} \\
	\end{tabular}
	\caption{ The mass contributions $r c_r(t)$ for $r$ and time for the Knowles (left) and Smoluchowski (right) models, respectively. In both plots, the solid black curve represents the average length of fibrils, $L(t)$. The parameters used in Knowles' and the Smoluchoswki model were the same as those used in Fig.~\ref{mass_plot_cgsa}, and the initial mass concentration was $m_{tot}=35 \mu$M. }
	\label{distro_plot}
	\end{center}
	\end{figure*}
%%%%%%%%%%%%%%%%%%%%%%%%%%%%%%%%%%%%%%%%

To illustrate these effects, we plot in Fig.~\ref{numb_plot}(c) and (d) the average lengths of aggregates, $L(t)$, i.e. Eq.~(\ref{length}), for both the Smoluchowski and Knowles' model using the parameters that best fit $M(t)$ to the Curli ThT data. The plots illustrate the fact that in both models the higher the initial mass concentration, the longer the fibrils may initially grow. At some time $t$ the average length of fibrils, $L(t)$, reaches a maximum value, and then starts to decrease as time moves forward. The curves are also more strongly peaked for greater initial mass concentration, where the peak broadens and decreases in maximum amplitude as the initial mass is decreased. For all initial mass concentrations, $m_{tot}$, considered in Knowles' model, the average length of fibrils, $L(t)$, decreases to only a fraction of its peak value as time progresses, eventually trending toward a common constant value once the system nears equilibrium. The peaks in the length curves must mean that the aggregates initially are few and over-shoot a stable average length, but as their numbers increase and monomers become scarce, fragmentation becomes more important, and their average lengths eventually approach a stable value near equilibrium. However, for the Smoluchowski model, clearly the effect of fragmentation is weakened at large times and the average aggregate length of fibrils, $L(t)$, does not decrease as significantly as in Knowles' model . Additionally, the average fibril lengths in the Smoluchowski model are longer than in Knowles' model, particularly at larger time scales. As is known, the fragmentation reactions can speed up the formation of small aggregates in both models for early times, but the possibility for aggregates of any size to merge into larger ones in the Smoluchowski model seems to counter-balance the breaking effect at greater times. To summarize, we observe for large times, Knowles' model  predicts that more and more aggregates will form, and the lengths of these aggregates get shorter and shorter. However, in the Smoluchowski model, the number of aggregates flattens out, as does the average lengths of these aggregates. 

For a better look at the predicted size distribution of the Curli fibrils, a contour plot is presented in Fig.~\ref{distro_plot} where each mass contribution $r c_{r}(t)$ is plotted against $r$ and time for both models using the initial concentration of 35$\mu$M for the Curli fibrils. In Fig.~\ref{distro_plot}(a), the fibril mass, $M(t)$, for Knowles' model at early times is initially spread out over many clusters sizes, but for longer times the overall fibril mass contributions become highly clustered around the average length value, $\langle L \rangle \approx$ 300, thus fluctuations about the average length are relatively modest. In Fig.~\ref{distro_plot}(b), the overall system size in the Smoluchowski model is significantly larger than for Knowles' model, where at long times the fibril mass, $M(t)$, in the Smoluchowski model is spread out amongst many cluster sizes, and that the fluctuation in mass (not shown) is quite large when compared to Knowles' model. Therefore, for the parameters used in the fit, fragmentation in the Smoluchowski model does not play as a significant role at later times as in Knowles' model, and that many cluster sizes contribute to the total fibril mass as compared to Knowles' model. The differences must attribute to the possibility of aggregates of any size to merge. Since the size of aggregates plays an important role in their toxicity, it is important for the interpretation and treatment to know which model, if any, describes the correct pathway. For this analysis, size-dependent kernels become necessary since the rates allow for determination of the properties of the aggregates that can be measured. The question then is which aggregation and fragmentation kernels should used in the Smoluchowski kinetics predictions for the average lengths, $L(t)$. 

\subsection{Size-dependent kernel results}
%%%%%%%%%%%%%%%%%%%%%%%%%%%%%%%%%%%%%%%% 
	\begin{figure*}[t]
	\begin{center}
	\includegraphics[width = 365 pt]{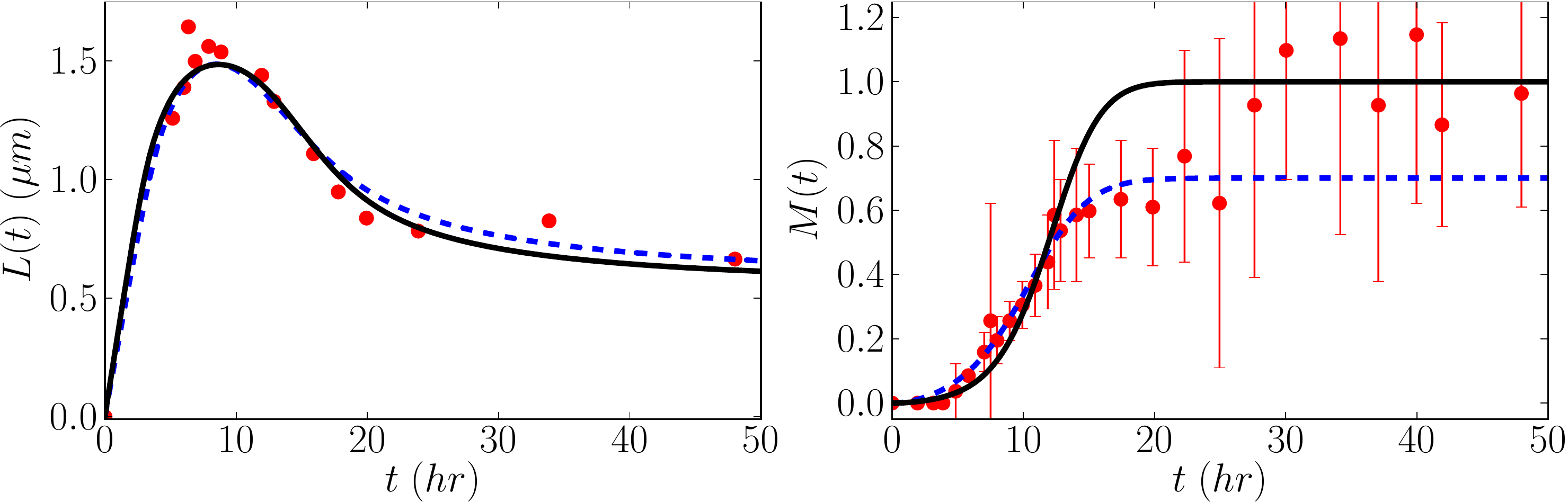} 
	\caption{ The comparison of $L(t)$ (left) to the experimentally determined average lengths of $\beta$-LAC fibrils. The data points are red dots. $M(t)$ (right) is plotted for the $\beta$-LAC fibrils. The fit parameters used for the black solid curve in both plots were $k_P =4.71\cdot10^2$~M$^{-1}$s$^{-1}$, $k_M = 1.74 \cdot 10^{-16}$ M$^{-1}$ s$^{-1}$, $k_n = 8.86$~M$^{-1}$ s$^{-1}$, and $D_f = 1.34$.  For the dashed blue curves in both plots, the parameters used were $k_P =8.16\cdot10^2$~s$^{-1}$, $k_M = 1.15 \cdot 10^{-16}$~M$^{-1}$ s$^{-1}$, $k_n = 1.8$~M$^{-1}$ s$^{-1}$, and $D_f = 1.35$. Additionally in both plots, $\lambda_b=8192$ and $n=4$ was used in Eq.~(\ref{frac_kernel}).  }
	\label{blac_lengths}
	\end{center}
	\end{figure*}
%%%%%%%%%%%%%%%%%%%%%%%%%%%%%%%%%%%%%%%%
We now show that the size-dependent aggregation and fragmentation kernels can be used to fit ThT and average length data. The average length of $\beta$-lactoglobulin ($\beta$-LAC) fibrils at acidic pH were measured using an AFM at different incubation times, where the initial mass of protein used was 10 g/L and $n_c=4$~\cite{Morbidelli2012}. The experiment showed that, starting from a monomeric solution, the proteins assembled into fibrils and steadily grew longer until peaking at 1.6 $\mu$m in length at 8 hours. The fibrils then steadily declined in length-- the average length was about 0.8$\mu$m at 20 hours. After 20 hours it appeared that the fibrils were only slightly shrinking in average length, though there were only a few measurements made after 20 hours. The experimental results for $L(t)$ for the $\beta$-LAC fibrils are plotted in Fig.~\ref{blac_lengths}. In order to compare the model predictions of $L(t)$ with the AFM results, the measured lengths must be converted to number of proteins, $r$, in the fibrils. A simple conversion is $r = 2 \times L / d$, where $L$ is the measured value of $L$ in nanometers. For A$\beta$, $d\approx 0.62$nm~\cite{Collins2004,Morbidelli2012}, here $d=1.5$ nm was used as a rough estimate of the distance between to two neighboring sheet-conformed $\beta$-LAC proteins in a fibril. This value of $d$ also helps facilitate shorter numerical computation times. Additionally, the factor of two refers to the assumption that two identical protofibrils that make up a fibril are aligned in register. 

%%%%%%%%%%%%%%%%%%%%%%%%%%%%%%%%%%%%%%%%
	\begin{figure}[t]
	\begin{center}
	\vspace{0.6 cm}
	\includegraphics[width = 245 pt ]{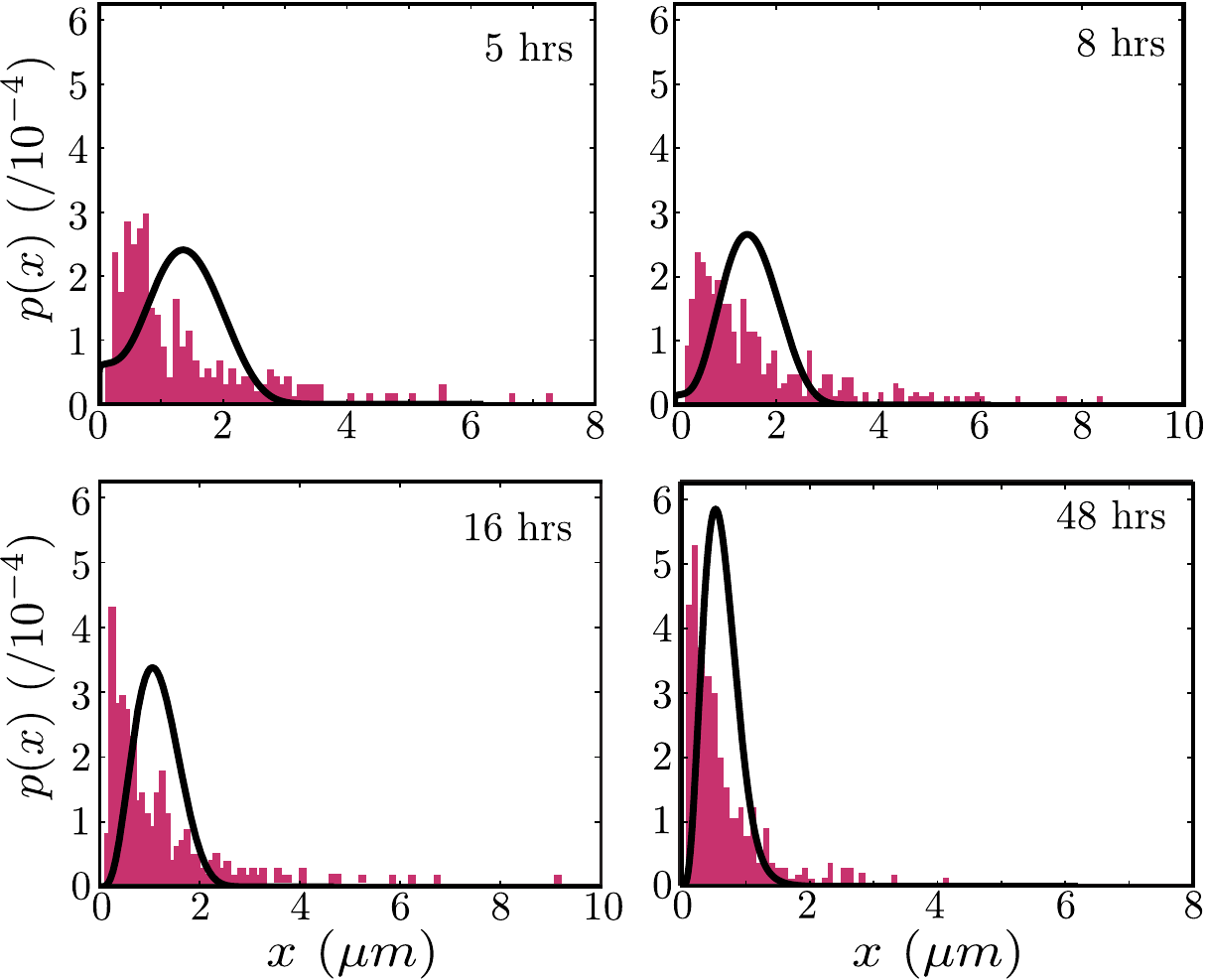}
	\caption{Predicted length distributions at 5, 8, 16, and 48 hours. In each plot, the shaded columns represent the AFM data with histogram bin widths of 10 nm, while the solid black curves represents the theoretical predictions. The parameters used were those that correspond with the solid black curve in Fig.~\ref{blac_lengths}.}
	\label{length_distro}
	\end{center}
	\end{figure}
%%%%%%%%%%%%%%%%%%%%%%%%%%%%%%%%%%%%%%%%

The average length of fibrils, $L(t)$, was computed by solving Eq.~(\ref{smoluchowski}) together with Eqs.~(\ref{sphere_kernel}) and (\ref{frac_kernel}). The comparison of $M(t)$ and $L(t)$ with the $\beta$-LAC mass and length measurements in Fig.~\ref{blac_lengths} shows agreement between theory and experiment. The solid black curves in Fig.~\ref{blac_lengths} represent the theoretical prediction for the case where all ThT data points were used, while the dashed blue curves represent the theoretical fit where only ThT data points up to 25 hours were used. In Fig.~\ref{length_distro}, the theoretical results are compared with the length distribution of the fibrils in the AFM experiment at various times. The theoretical distribution used to compare with the results in Fig.~\ref{length_distro} was $P_r(t) = c_r(t) / \sum_{r=n_c}^{\infty} 2 c_r(t)$ where the factor of two was introduced because we assumed that each fibril contained two filaments. In addition to the adjustable constants $k_P$, $k_M$, and $k_n$, the fractal dimension $D_f$ was found to be $D_f \approx 1.35$. As reported earlier, fractal dimensions for other non-amyloid protein aggregates have been found to be $D_f \approx 2.0-2.7$~\cite{Gimel1994, Hagiwara1997, Arosio2012}. For amyloid fibrils, the aggregates are long, and rod-like in shape, thus we might expect that $D_f \sim 1$.  

Arosio et al.~\cite{Morbidelli2012}, and others\cite{Collins2004}, have showed that a model similar to the one used by Knowles et al.~\cite{Knowles2009} needs an adjustment to the constant fragmentation rates to fit the same data for $\beta$-LAC fibrils. If the rate of breaking, $k_-$, like in Eq.~(\ref{knowles}), is assumed to possess a size dependence $\sim r^{\lambda}$, where $r$ is the size of the aggregate, then using $\lambda=3$ allows Arosio's model accurately predict the average lengths of the fibrils. We have shown that the aggregation and fragmentation kernels derived from diffusion theory can accurately predict the ThT and AFM data for the $\beta$-LAC fibrils. 

\section{Generalized kinetics}
\label{sec:gen_kinetics}

Now, we return to the scenario where oligomers may reach metastable states before becoming fibrils. As discussed in the section ``Critical Nuclei'', some of the small micelle-like aggregates could undergo a conformational transition to B-type nuclei when their sizes reach the critical nucleus. B-type aggregates could then grow into fibrils that could be long and contain large amounts of $\beta$-structure. Since both A-type and B-type aggregates are kinetically active at the same time, they may compete for available monomers~\cite{wattis1999} in order to grow larger. A more general scenario, where small micelle-like aggregates containing $n_c$ number of proteins may form, is described by the reactions
\begin{eqnarray}
\label{micelle_reaction}
C_1+ C_1 &\underset{k_-^A}{\overset{k_+^A}{\rightleftharpoons}}& C_{2}^A \\ 
C_1+ C_2^A &\underset{k_-^A}{\overset{k_+^A}{\rightleftharpoons}}& C_{3}^A \nonumber \\ 
&\vdots& \nonumber \\ 
C_1+ C_{n_c-1}^A &\underset{k_-^A}{\overset{k_+^A}{\rightleftharpoons}}& C_{n_c}^A \nonumber
\end{eqnarray}
where $k_+^A$ is the monomer addition rate constant, and $k_-^A$ is the monomer subtraction rate constant. For simplicity it was assumed that all forward rates have the same value, and likewise for the backwards rates.  

The mass-action equations governing the evolution of the concentrations for $A$ and $B$ types can be written using Becker-D{\"o}ring or Smoluchowski kinetics. For example, using the Smoluchowski kinetics for the A- and B-type aggregates, the mass-action equations governing the kinetics can be written as 
\begin{eqnarray}
\label{a_type}
\frac{d c_{r}^A(t)}{dt} &=& W_{r-1,1}^A(t) - W_{r,1}^A(t) - k_{n} c_{n_c}^A(t) \delta_{r,n_c} \\ 
\label{b_type}
\frac{d c_{r}^B(t)}{dt} &=& \frac{1}{2} \sum_{s=1}^{r-1} W_{s,r-s}^B(t) - \sum_{s=1}^{\infty} W_{r,s}^B(t) \nonumber \\
 &&\! {}+ k_{n} c_{n_c}^A(t) \delta_{r,n_c}  \\
\label{mon_term}
\frac{d c_{1}(t)}{dt} &=& - \sum_{r=2}^{n_c} r \frac{ d c_{r}^{A}(t) }{dt} - \sum_{r=n_c}^{\infty} r \frac{ d c_{r}^{B}(t) }{dt}
\end{eqnarray}
where the fluxes $W_{r,s}$ were defined in Eq.~(\ref{smolflux2}) and we assumed that the A-type aggregates cannot grow larger than the critical size $n_c$. B-type aggregates must be at least the size of the critical nucleus. The third term in Eqs.~(\ref{a_type}) and (\ref{b_type}) represents the loss of A-type and the gain of B-type aggregates of size $n_c$, respectively. No longer is the nucleus production term in Eq.~(\ref{b_type}) simply proportional to the monomer concentration raised to the power $n_c$, as it was in Eq.~(\ref{nuke_eq2}) when we assumed the A-type micelles rapidly reached equilibrium. Eqs.~(\ref{a_type}-\ref{mon_term}) are easily solved numerically. 

%%%%%%%%%%%%%%%%%%%%%%%%%%%%%%%%%%%%%%%%
	\begin{figure}[t]
	\begin{center}
	\vspace{0.6 cm}
	\includegraphics[width =245 pt ]{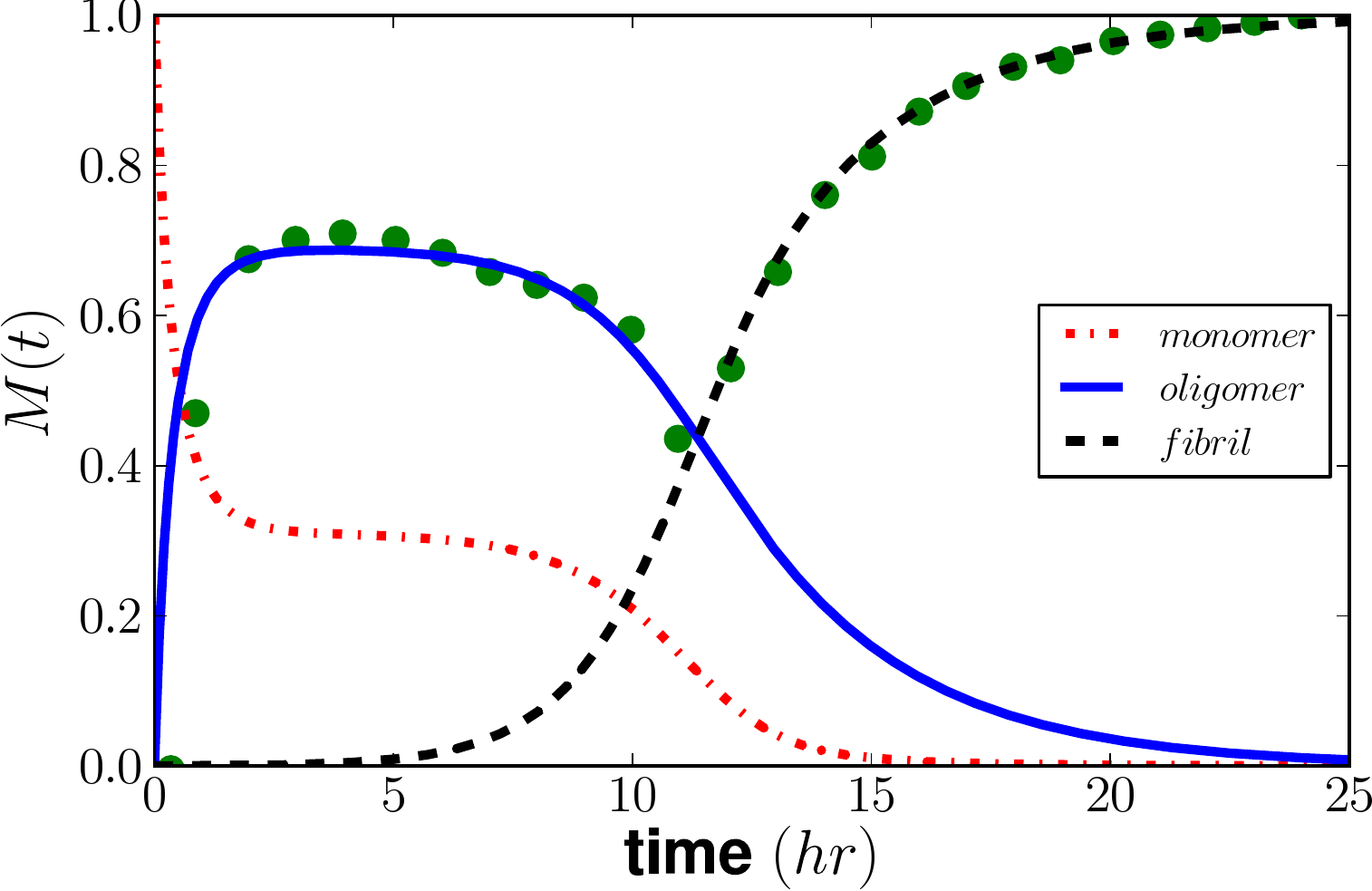}
	\caption{Predicted masses of $A$ (solid blue line) and $B$ (dashed black line) type aggregates. Monomers are also plotted (dashed-dotted red line). The concentrations of $B$-type aggregates were determined using Knowles' model, i.e. Eq.~(\ref{knowles}). Green circles are the selected FlAsH fluorescence data points from Lee et al.~\cite{Kelly2011}. The $A$-type oligomer for A$\beta$(1-40) was assumed to be a dimer. The $A$-type oligomers may then undergo a conformational transition to $B$-type critical nuclei. $B$-type critical nuclei may proceed to grow into amyloid fibrils. Parameters used in the fit were $m_{tot}=10 ~\mu$M, $k_+^A = 269$~M$^{-1}$ s$^{-1}$, $k_-^A=0.00015$~s$^{-1}$, $k_{+}^B=1.3\cdot 10^{4}$~M$^{-1}$ s$^{-1}$, $k_{-}^{B}=3.87 \cdot 10^{-6}$ s$^{-1}$, $k_n=2.64\cdot10^{-6}$~M$^{-1}$ s$^{-1}$.}
	\label{kelly_plot}
	\end{center}
	\end{figure}
%%%%%%%%%%%%%%%%%%%%%%%%%%%%%%%%%%%%%%%%

An example calculation of the mass in aggregates is shown in Fig.~\ref{kelly_plot} based on Eq.~(\ref{knowles}) with Eqs.~(\ref{a_type}) and (\ref{mon_term}). The A-type oligomers may grow to the size of the critical nucleus, $n_c$, and convert into sheet, or shrink by monomer subtraction. The plot clearly shows an oligomer phase, followed by a transition phase where the oligomers either dissociate, or convert to $B$-type nuclei. Eventually, the aggregates mostly become $B$-type fibrils. These results should be compared with the experimental results obtained by Lee et al.~\cite{Kelly2011}, which show similar FlAsH ThT florescence curves for the masses in oligomers and fibrils. Additionally, the curves in Fig.~\ref{kelly_plot} seem to agree with the theoretical results obtained by Lee~\cite{lee}.

\section{Discussion and Conclusion}
In this article, we studied the kinetics of protein aggregation using several mass-action models that describe the evolution of aggregate concentrations $c_r(t)$. The NP and Knowles' models have been well-studied in the literature. However, the Smoluchowski model for protein aggregation has received very little attention. Knowles' model, and our proposed Smoluchowski model for protein aggregation, both have been shown to fit ThT curves reliably, where each model has the feature that aggregate breaking can act as a feedback loop, where the fragments can act as seeds, and could range in size from the critical nucleus, or larger. However, when turning attention to the model predictions for the length distribution of aggregates, $L(t)$, the models predict very different results when compared to each other. 

The average lengths of aggregates in the Smoluchowski and Knowles models both exhibit a maximum average length at some time $t$, which then decreases at later times. The Smoluchowski model predicts that the aggregates will shrink down in size at later times, but not nearly as quickly as those aggregates described by Knowles' model. Thus, the average aggregate size described by the Smoluchowski model with constant kernels is found to be larger than that predicted by Knowles' model. But more importantly, kinetic models with constant kernels cannot be made to fit the length distribution data obtained for $\beta$-LAC fibrils using AFM. On the other hand, we show that by introducing length-dependent kernels to the Smoluchowski model yields results in good agreement with experimental data. 

Additionally, we also show that a kinetics model that allows for oligomer concentrations of aggregates smaller than $n_c$ can describe the conformational conversation of a micelle to a more ordered nucleus, which may then grow into fibrils. The theoretical results obtained seem to agree qualitatively with the recent FlAsH fluorescence results obtained by Lee et al.~\cite{Kelly2011} in which a metastable oligomer phase seems to precede the elongation of fibrils.

\section*{Acknowledgments}
We would like to thank Profs. Frank Ferrone and Brigita Urbanc for useful discussions. JSS would like to thank Drs. Renan Cabrera and Tom Williams for helpful discussions.

\bibliography{meta}

\end{document}